\documentstyle[12pt]{article}
\begin{document}
\baselineskip 17pt plus 2pt

\hspace{9cm}
          \begin{large}
BRX-TH-342\\
          \end{large}
\vspace{0.6cm}
           \begin{center}
           \begin{Large}
           \begin{bf}
The Wave Function of a Collapsing Star  \\
and Quantization Conditions \\
            \end{bf}
            \end{Large}
\vspace{1.5cm}
YOAV PELEG\footnote{This work is supported by the Fishbach Fellowship
92/3 and by the NSF grant PHY 88-04561.} \\
\vspace{0.7cm}
Physics Department, Brandeis University, Waltham, MA 02254-9110\\
\vspace{0.4cm}
{\em Peleg@Brandeis}\\
\vspace{1cm}
           \begin{large}
ABSTRACT\\
            \end{large}
            \end{center}
\vspace{0.5cm}

A very simple minisuperspace describing the Oppenheimer-Snyder collapsing star
is found. The semiclassical wave function of that model turn out to
describe a bound state.
For fixed initial radius of the collapsing star,
the corresponding Bohr-Sommerfeld
quantization condition implies mass quantization.
An extension of this model, and some consequences, are considered.

\newpage

The simplest model of a spherically symmetric homogenous dust
collapsing star is that of Oppenheimer and Snyder [1].
In that model one connects continuously
the Schwarzschild interior to the interior region, which is a
slice of a Friedmann universe.

To construct the (mini) superspace of that model one needs the
relevant Lagrangian and Hamiltonian. Generally one does not use the
Lagrangian description of a dust distribution, because its density
$ \rho (x) $ , is not a dynamical variable. Instead one uses the
energy momentum tensor, $ T_{\mu \nu} = \rho U_{\mu} U_{\nu} $ ,
where $ U_{\mu} $ is the
four-velocity of the matter particles. The Einstein equations are
$ G_{\mu \nu} = 8 \pi T_{\mu \nu} $ . In [2] it was shown how one
can define a matter dust Lagrangian that will lead to those equations
of motion.

We will use geometrical units: $ G=c=1 $ and signature $ (-+++)$.
The interior region of the Oppenheimer-Snyder model is described by
the Friedmann line element,
   \begin{equation}
d s^{2} = - N^{2}(t)d t^{2} + a^{2}(t)[d{\chi}^{2} + sin^{2}\chi
d{\Omega}^{2}_{2} ]
   \end{equation}
The range of $ \chi $ is $ 0 \leq \chi \leq {\chi}_{0} $ , where $ {\chi}_{0}
< \pi/2 $ . At $ \chi = {\chi}_{0} $ the interior is matched to the exterior
Schwarzschild solution.
At least classically, the energy momentum tensor must be conserved,
 $ \nabla_{\mu} T^{\mu \nu} = 0 $ . In the case of a Friedmann
universe, this implies that $ \rho = \rho_{0}/a^{3} $ , where $ \rho_{0} $
is a constant determine by the initial conditions. In the collapse case
one gets $ \rho_{0} = 3a_{0}/8\pi $ , where $ a_{0} $ is the radius of the
Friedmann ball at the beginning of the collapse.
So the density of the star is
   \begin{equation}
   \rho = \left\{ \begin{array}{ccc}
   3 a_{0} / 8\pi a^{3}(t) & , & \chi \leq \chi_{0} \mbox{ } (r \leq r_{s})
\nonumber \\
                0          & , & \chi > \chi_{0} \mbox{ }(r > r_{s})
              \end{array} \right.
   \end{equation}
where $ r_{s} $ is the surface radius of the collapsing star.

To get the matter Lagrangian one defines the ``weight one density"
$ \tilde{\rho} \equiv \sqrt{-g} \rho $ [2] (which is the generalization
of a delta function). Then the matter Lagrangian is
$ L_{M} = -8\pi \int \tilde{\rho} g^{\mu \nu} U_{\mu} U_{\nu} d^{3}x $ .
The four-velocity satisfies (classically)
$ U^{\mu} U_{\mu} = -1 $ , so from (1) and (2) we get
   \begin{equation}
L_{M} = 32{\pi}^{2} \int_{0}^{{\chi}_{0}} sin^{2} \chi d\chi Na^{3} \rho ~.
   \end{equation}
The gravitational Lagrangian may be split into its interior and exterior
parts,
   \begin{equation}
L_{G} = 4\pi \int_{0}^{{\chi}_{0}} sin^{2}\chi d\chi \left[
\frac{3a}{N} {\dot{a}}^{2} - 3Na \right] + \int_{r \geq r_{s}}
\sqrt{-g} R d^{3}x ~.
   \end{equation}
The total Lagrangian is then
   \begin{equation}
L = L_{G} + L_{M} = 4\pi \int_{0}^{\chi_{0}}sin^{2}\chi d\chi
\left[ \frac{a {\dot{a}}^{2}}{N}
-N(a - a_{0})  \right] + \int_{r\geq r_{s}} \sqrt{-g}R d^{3}x
   \end{equation}
The matching conditions are [1]
   \begin{eqnarray}
M &=& \frac{1}{2} a_{0} sin^{3}{\chi}_{0} \nonumber \\
R_{0} &=& a_{0} sin{\chi}_{0}
   \end{eqnarray}
where $ R_{0} $ is the radius of the surface of the star at the beginning
of the collapse, and $ M $ is the mass of the Schwarzschild solution.
As one can see from (6),  $ {\chi}_{0} $ is
determined by the initial condition: $ 2M/R_{0} = sin^{2}{\chi}_{0} $ .
For a reasonable cosmological initial condition we have $ R_{0} >> M $ so:
$ {\chi}_{0} <<  1 $ . In that case the integral over the slice
of $ S^{3} $ is $ 4\pi{\chi_{0}}^{3}/3 $ , and
   \begin{equation}
M \approx \frac{1}{2} R_{0} \chi_{0}^{2} ~.
   \end{equation}
Hence
   \begin{equation}
L = 4\pi \chi_{0}^{3} \left[ \frac{a {\dot{a}}^{2}}{N}
-N(a - a_{0})  \right] + \int_{r\geq r_{s}} \sqrt{-g}R d^{3}x ~.
   \end{equation}

The classical solution of (5) (or (8)) is easily obtained\footnote{
Note that for our dust, the only variables in (5) (or (8)) are
the metric components ($a,N$) since the the matter field $\rho$ is
determined by $a$; the two corresponding Euler-Lagrange equations
are of course equivalent
to the Einstein ones with the dust source.}: For $ r \geq r_{s} $ ,
we have the Schwarzschild solution. For $ r \leq r_{s} $ $ (
\chi \leq \chi_{0} ) $, it is convenient to choose the gauge for
which $ N(\eta) = a(\eta) $. Then the solution for which $ a(\eta=0)=
a_{0} $ and $ \dot{a}(\eta=0)=0 $ , is
   \begin{equation}
a(\eta) = a_{0}(1+cos\eta)/2 ~.
   \end{equation}

Our strategy is this: (5) (or (8)) is our model. This is our starting
point. Classically it describes the Oppenheimer-Snyder collapse process;
we will find the semiclassical wave function of this universe. This
approach is consistent with the analysis of [3], in which
a general (inhomogeneous) distribution of spherically symetric dust
matter is considered. The field describing the matter (the dust) is
eliminated by gauge conditions and one ends up with only gravitational
degrees of freedom.

Usually one starts from a known matter distribution and tries to find
the metric. Here we go in the opposite direction: we first determine
the geometry (the metric), and then find the matter distribution.
Classically both directions are equivalent. Semiclassically we can use
our model, because we need only the classical action. Higher quantum
orders will require a modification of this model\footnote{Because for
example $ U^{\mu}U_{\mu} \neq -1 $, and we cannot get (3).}.
We will see that the WKB wave function of that universe describes a
{\em bound state}, so we will get quantization conditions.

One can use the same formulation to study closed Friedmann
minisuperspace models filled with dust (and/or radiation) [4].
In that case $ 0 \leq \chi \leq 2\pi $ , and there is no ``outside"
Schwarzschild region.

The semiclassical wave function\footnote{We work in the Lorentzian
section but one can get exaclty the same results working in the
Euclidian section. This is true also for the standard minisuperspace
models, and one may argue that it should be the same in the full
superspace.} of this collapsing star is
   \begin{equation}
\psi_{_{WKB}}(t) = A(t)exp[iS_{Class.}(0,t)/\hbar]
   \end{equation}
where
   \begin{equation}
S_{Class.}(0,t) = 4\pi \chi_{0}^{3} \int_{0}^{t} L(t')dt' +
             \int_{0}^{t} dt' \int_{r\geq r_{s}} \sqrt{-g}R_{Schw.}d^{3}x~.
   \end{equation}
Because $ R_{Schw.}=0 $ , only the first term in the r.h.s of (11) will
contribute to the semiclassical wave function. We can calculate (11)
very easily in the gauge $ N(\eta) = a(\eta) $ , for which
   \begin{equation}
\psi_{_{WKB}} = C {\left( \pi \chi_{0}^{3} a_{0}^{2}(\eta - \frac{1}{2}
sin{2\eta} ) \right) }^{-1/2}
exp \left( \frac{i}{\hbar} \pi \chi_{0}^{3} a_{0}^{2}(\eta - \frac{1}{2}
          sin{2\eta} ) \right) ~.
   \end{equation}
One can easily see that (12) is the WKB solution of the one-dimensional
Schrodinger equation
   \begin{equation}
\left[ -\hbar^{2} \frac{d^{2}}{d a^{2}} + V(a) \right] \psi(a) = 0
   \end{equation}
where
   \begin{equation}
V(a) = {(8\pi \chi_{0}^{3})}^{2} a (a-a_{0}) \mbox{  } , \mbox{  }
0 \leq a \leq a_{0}~.
   \end{equation}
The semiclassical solution of (13), (14) is
   \begin{equation}
\psi_{_{WKB}}(a) = C (p(a))^{-1/2} exp[i p(a)/\hbar]
   \end{equation}
where
   \begin{eqnarray}
p(a) &=& \int_{a_{0}}^{a} \sqrt{E - V(a')} da' = \int_{a_{0}}^{a}
\sqrt{|V(a')|} da' \\
     &=& \pi \chi_{0}^{3} \left[ 2(2a-a_{0})(a_{0}a-a^{2})^{1/2}
         - a_{0}^{2}\left( arcsin(1-2a/a_{0}) + \frac{\pi}{2} \right)
          \right] \nonumber
   \end{eqnarray}
Using (9), one can see that (12) and (15) are identical.

The Schrodinger equation (13),(14) is just the Wheeler-DeWitt equation
of this simple model [5,6,7]. To see this, we go back to (8) and construct
the Hamiltonian in the standard way. One should remember that only the
first term in the r.h.s of (8) contributes to the semiclassical dynamics.
The Hamiltonian is
   \begin{equation}
H = \dot{a} P_{a} - L
   \end{equation}
where $ P_{a} = \partial L / \partial \dot{a} = 8 \pi \chi_{0}^{3}
a\dot{a}/N $ , namely
   \begin{equation}
H = N \left[ \frac{1}{16\pi\chi_{0}^{3} a} P^{2}_{a} +
             4\pi \chi_{0}^{3} (a-a_{0}) \right]
   \end{equation}
The Wheeler-DeWitt equation is the quantum version of the classical
Hamiltonian constraint, $ \partial H / \partial N = 0 $ ,
in the coordinate representation $ |\psi> = \psi(a) $ and
$ P_{a} \rightarrow -i\hbar \partial / \partial a $ . Using
(18) we get
   \begin{equation}
\frac{1}{16 \pi {\chi_{0}}^{3} a} \left[ -\hbar^{2} \frac{d^{2}}{d a^{2}} +
V(a) \right] \psi(a) = 0
   \end{equation}
where the potential is (14). So (13) is (as we expected) the Wheeler-
DeWitt equation of this model.

We see from (14) that there are two turning points, at $ a=0 $ and
$ a=a_{0} $ , between which there is a {\em bound state}\footnote{
The wave function of this bound state is of course real
   \begin{eqnarray*}
\Psi_{WKB}(a) = c {(p(a)}^{-1/2} ( e^{ip(a)/\hbar} + e^{-ip(a)/\hbar})
   \end{eqnarray*}
which corresponds to the Hartle-Hawking wave function [7].}.

Consequently there will be {\em quantization conditions}.
In our semiclassical limit (in which General Relativity should
be a good approximation) the Bohr-Sommerfeld quantization
condition says that
   \begin{equation}
\int_{0}^{a_{0}} \sqrt{|E - V(a)|} da = \int_{0}^{a_{0}} \sqrt{|V(a)|} da
=  (n + 1/2) \pi \hbar ~.
   \end{equation}
Let $ R_{0} $ be the initial radius of the collapsing star. Then (20) is
a quantization condition on the mass of the star. From (7), (14) and (20)
we get the mass quantization condition
   \begin{equation}
M(n) = \frac{1}{2\pi^{2}} {\left( \frac{l_{P}}{R_{0}} \right)}^{3} (n+1/2)^{2}
    M_{P}
   \end{equation}
where $ l_{P} $ and $ M_{P} $ are the Planck length and mass (in the
geometrical units $ l_{P} = M_{P} = \hbar^{1/2} $).

Before we consider some of the consequences of (21),
it is important to verify its extension. Is it model dependent,
and in what way?

In the Appendix we consider in details the addition of a conformally
invariant scalar field as a small perturbation. The resulting Wheeler-DeWitt
equation is (see (41),(42))
   \begin{equation}
\left[ -\hbar^{2} \frac{d^{2}}{da^{2}}+\tilde{V}_{m}(a) \right] \psi(a) = 0
   \end{equation}
where
   \begin{equation}
\tilde{V}_{m}(a) = {(8\pi \chi_{0}^{3})}^{2} \left[ a \left(a - a_{0} -
\frac{\alpha}{4\pi \chi_{0}^{3}} \right) - \frac{(m+1/2)\hbar }{4\pi
\chi_{0}^{3}} \right]
   \end{equation}
where $ \alpha $ is a real number, and $ m $ is an integer (it is the
quantum number of the scalar field inside the star).
Comparing (14) with (23), we see that small perturbation requires that
$ a_{0} >> \alpha / 4\pi \chi_{0}^{3} $ and $ n >> m $.
The classical equation of motion
that we get from (23) will be exactly like the unperturbed one, if we
replace  $ a_{0} \rightarrow b_{0} = a_{0} + \alpha / (4\pi \chi_{0}^{3})
$ . Now, one can easily see that for $ n >> m $, the quantization condition
   \begin{equation}
\int_{0}^{b_{0}} \sqrt{|\tilde{V}_{m}(a)|}da = (n+1/2) \pi \hbar
   \end{equation}
leads to the mass quantization condition
   \begin{equation}
M(n,m) \approx \frac{1}{2\pi^{2}}{\left( \frac{l_{P}}{R_{0}} \right)}^{3}
{\left( \sqrt{n+m+1} - \sqrt{m+1/2} \right)}^{4} M_{P}
   \end{equation}
which is slightly different from (21),
(in the ``limit": ``$ m + 1/2 \rightarrow 0 $" we of course recover (21)).

This example shows that one can get quantization condition in a more general
case than the simple Oppenheimer-Snyder model. But if,
for example, we make the same calculation for a minimally coupled massless
scalar field, we get the potential
   \begin{equation}
W(a) = {(8\pi \chi_{0}^{3})}^{2} \left[ a \left(a - a_{0} -
\frac{\alpha}{4\pi \chi_{0}^{3}} \right) - \frac{k^{2}}{4\pi
  \chi_{0}^{3}a^{2}} \right]
   \end{equation}
where $ k $ is a real number. But now the integral
   \begin{equation}
\int_{0}^{b_{0}} \sqrt{|W(a)|} da
   \end{equation}
diverge logarithmically,
and there are no quantization conditions! The problem is that
when $ a \rightarrow 0 $ , the potential (26) diverges.
This means that the perturbations are not small at $ a \rightarrow
0 $ , and we can no longer say that there is a collapse process.
The classical solutions of (26) do not describe a collapse! This seems to
contradict Price's results [8] that the Oppenheimer-Snyder model is stable.
However, one should remember that to get (23) or (26), we neglect a very
important condition, continuity at the surface of the star. So as a
matter of fact
(23) and (26) are incorrect. We can see this from the
following consideration:
from the no-hair theorems we know that for late times the
scalar field outside the horizon vanishes. From continuity (and the fact
that we consider only constant scalar field inside the star), it should vanish
also inside the star. In (26) $ k $ (and in (23) $m$ ) is the quantum number
of the scalar field inside the star. So if somehow (using the correct
continuity conditions) we were able to
get an effective one-dimensional Wheeler-DeWitt equation (like (22)
with (23) (or (27)))
then for late times $ k \rightarrow 0 $ ( $ m \rightarrow 0 $ ). Late
times corresponds to small $a$, so $k$ (and $m$) should be $a$-dependent.
In that case (27) will not diverge.
It is not easy to find the whole superspace with the boundary conditions
[9], and to solve the Wheeler-DeWitt equation. But if the
Oppenheimer-Snyder model is stable, then for small perturbations, one
should end up with an effective potential (such as we
got in the over-simplified model of the conformally invariant scalar
field\footnote{Probably the continuity conditions will not change (24) too
much, because anyway $\alpha $ and $ m $ are small and contribute a
finite correction in (24).} ) which is close to (14).
One is tempted to make the conjecture that
{\em the WKB wave function of a (general) collapsing star
describes a bound state, leading to quantization conditions.}  \\
But one should find first the WKB solution of the full Wheeler-DeWitt
equation, or at least the full linear perturbations, which is not an
easy task!

The reason that there are no quantization condition in the standard
cosmological model of Friedmann universe [7], is that in that
model the potential is (when there is a cosmological constant\footnote{
Otherwise the solution is not very interesting.}, $ \Lambda $):
   \begin{eqnarray*}
V(a) =   a^{2} - H^{2}a^{4}
   \end{eqnarray*}
where $ H^{2} =  \Lambda / 3 \geq 0 $.
This potential does not describe a bound state.

We end with some comments:

1) It can be easily seen that our quantization conditions do not effect
the classical
collapse process, because of the correspondence limit\footnote{For our
sun, for example,
$ n \sim 10^{100} $ !}
   \begin{equation}
\frac{\Delta M}{M} = \frac{M(n) - M(n-1)}{M(n)} \sim \frac{1}{n} << 1~.
   \end{equation}

2) We obtained a ``strange" quantization condition. It depends strongly
on the initial
value $R_{0}$. As a matter of fact if $R_{0}$ is not much greater than
$M$, one should replace $ \chi_{0}^{3} $ in (14) (or (24)) with
$ 3(\chi_{0}/2 - sin(2\chi_{0})/4) $. Then the quantization condition
(21) will be different, so the exact $ M $ is a more complicated function
of $ R_{0} $. The strange feature of our quantization condition is that
the mass quanta are generally much smaller then the Planck mass,
a more ``reasonable" quantization condition should not
involve $ R_{0} $ in an essential way\footnote{For example
$ M \sim f(n) M_{P} $, such that $f$ is a function of $n$ but not of
$R_{0}$ .}.

3) To see if the Hawking process can be affected by the quantization
condition, one should compare the energy of a radiated particle to the
mass difference $ \Delta M $ . The energy of a radiated particle is:
   \begin{equation}
E_{r.p} \sim \frac{1}{2}k_{B}T = \frac{M_{P}^{2}}{16\pi M} ~.
   \end{equation}
The Hawking process should not be affected, if $ E_{r.p} >> \Delta M $ .
In that case one can consider the mass as continuous. In our case this
leads to $ n << R_{0} / l_{P} $,  which (in our geometrical units) means that
   \begin{equation}
M << R_{0} ~,
   \end{equation}
consistent with reasonable cosmological initial conditions. So
our quantization condition should not affect the Hawking process.
It is interesting to notice that the more ``reasonable" quantization
condition, $ M(n) = f(n) M_{P} $, {\em would} (very much) affect the Hawking
process, because (in that case) $ E_{r.p} >> \Delta M $ leads to
$ M << M_{P} $ . So perhaps our ``strange" quantization condition is
not so strange?

4) Can strong quantum effects drastically change our result? Quantum
effects should be dominant when $ a \rightarrow 0 $ . But if those effects
should save us from the singularity, or make it less singular, then
they must (drastically) increase the
potential (near $ a = 0 $ ). This means that we will have an even stronger
binding, and quantization conditions should remain.

Finally, it is very important to study
consistently the effect of  small perturbations around this minisuperspace.

\vspace{1cm}
{\bf Acknowledgment}\\
I would like to thank Stanley Deser for very helpful discussions.

\vspace{2cm}

\appendix
\section{Addition of a conformally invariant scalar field:}
In this Appendix we consider the addition of a conformally
invariant scalar field, $ \varphi (t,\vec{x}) $ .
For $ \chi \leq \chi_{0} $ ( $ r \leq r_{s} $ ) we take the scalar
field to be only time dependent, $ \varphi(t) $ (like in the standard
minisuperspace models [7]). But for $ r > r_{s} $ the scalar field must
be a function of the radial coordinate too, $ \varphi (t,r) $ .
The Lagrangian (8) is now:
   \begin{eqnarray}
L &=& L_{G} + L_{M} + L_{\phi} = \nonumber \\
  &=& 4\pi \chi_{0}^{3} \left[ \frac{a\dot{a}^{2}}{N} -
      \frac{a{\dot{\phi}}^{2}}{N} - N (a - a_{0} + \phi^{2}/a) \right] \\
     &+&  \int_{r>r_{s}} \sqrt{-g}\left[ R(1-\varphi^{2}/6) -
         {(\nabla \varphi)}^{2} \right] d^{3}x \nonumber
   \end{eqnarray}
where $ \phi(t) = a(t)\varphi(t) $ .
We consider the scalar field as a small perturbation to the
Oppenheimer-Snyder model. One can use ``generalized Novikov
coordinates" [10], for $ r > r_{s} $ , whose line element is:
   \begin{equation}
ds^{2} = - N^{2}(R,\xi) d\xi^{2} + F^{2}(R,\xi) dR^{2} + r^{2}(R,\xi)
d\Omega_{2}^{2} ~.
   \end{equation}
The Hamiltonian related to (31) is
   \begin{eqnarray}
H &=& H_{1} + H_{2} \nonumber \\
H_{1} &=& N\left[ \frac{1}{16\pi\chi_{0}^{3}a}(P_{a}^{2} - P_{\phi}^{2})
+ 4\pi\chi_{0}^{3}(a - a_{0} - \phi^{2}/a) \right] \nonumber \\
H_{2} &=& \int_{R_{0}}^{\infty} N(R,\xi) \left[ \frac{1}{32\pi}
   \left(-\frac{2}{Fr^{2}} P_{\varphi}^{2} + \frac{F}{r^{2}} P_{F}^{2}
   + \frac{1}{r} P_{F}P_{r} \right) \right. \nonumber \\
   & & \left.  + \left( ^{(3)}R - \frac{1}{F^{2}} {\left( \frac{
       \partial \varphi}{\partial R} \right)}^{2} - \frac{ ^{(4)}R
       \varphi}{6}  \right)Fr^{2} \right] dR
   \end{eqnarray}
where $ P_{\phi}=\partial L / \partial \dot{\phi} ~,~ P_{F} = \partial L
/ \partial \dot{F} ~,~ P_{r}=\partial L
/ \partial \dot{r} $ and $ P_{\varphi}=\partial / \partial \dot{
\varphi} $ .
The corresponding Wheeler-DeWitt equation is
   \begin{equation}
\left(\frac{\partial H}{\partial N}\right) |\Psi> = ( \hat{A}_{1} +
\hat{A}_{2} ) |\Psi> = 0
   \end{equation}
where
   \begin{eqnarray}
\hat{A}_{1} &=& \frac{1}{16\pi\chi_{0}^{3}a}\left[ \hat{P}_{a}^{2} -
\hat{P}_{\phi}^{2}
+ {(8\pi\chi_{0}^{3})}^{2}[a(a-a_{0}) - \phi^{2}] \right] \\
\hat{A}_{2} &=& \int_{R_{0}}^{\infty} \left[ \frac{1}{32\pi}
   \left(-\frac{2}{Fr^{2}} \hat{P}_{\varphi}^{2} + \frac{F}{r^{2}}
\hat{P}_{F}^{2}
   + \frac{1}{r} \hat{P}_{F}\hat{P}_{r} \right) \right. \nonumber \\
   & & \left.  + \left(^{(3)}R - \frac{1}{F^{2}} {\left( \frac{
       \partial \varphi}{\partial R} \right)}^{2} - \frac{
       ^{(4)}R \varphi}{6} \right)Fr^{2} \right] dR
   \end{eqnarray}
and $ \hat{P}_{\eta} = -i\hbar \partial / \partial \eta ~,~
\eta \equiv (a,\phi,F(R),r(R),\varphi(R)) $ .
We see that $ \hat{A}_{1} $ is a second order differential operator acting
on a two-dimensional space. But $ \hat{A}_{2} $ acts on an infinite
dimensional space (functional space). It is a very complicated space,
with boundary conditions result from gauge fixing and continuity
conditions at $ R = R_{0} $.
Because we are not going to solve the
Wheeler-DeWitt equation in that region anyway , we will not consider all the
details. Let $ \vec{z} $ be a coordinates in that space. And
let $ x \equiv a $ , $ y \equiv \phi $ . Then we make a separation of
variables\footnote{The coordinates $ x , y $ and $ \vec{z} $ , are
not independent. They are related by the continuity conditions.
We will work in this simplified model and then check if our results
are ``reasonable".} :
   \begin{equation}
\Psi(x,y,\vec{z}) = \Psi_{1}(x,y)\Psi_{2}(\vec{z})
   \end{equation}
then we get from (34) and (37)
   \begin{eqnarray}
\hat{A}_{1} \Psi_{1}(x,y) &=& \alpha \Psi_{1}(x,y) \\
\hat{A}_{2} \Psi_{2}(\vec{z}) &=& -\alpha \Psi_{2}(\vec{z})
   \end{eqnarray}
where $ \alpha $ is a (positive) number. From (39) and (42) we get:
   \begin{equation}
\left\{ -\hbar^{2} \frac{\partial ^{2}}{\partial x^{2}} +
       \hbar^{2} \frac{\partial ^{2}}{\partial y^{2}}  +
       {(8\pi\chi_{0}^{3})}^{2} \left[ x\left(x-a_{0}-
       \frac{\alpha}{4\pi\chi_{0}^{3}}\right) - y^{2} \right] \right\}
       \Psi_{1}(x,y) = 0 ~.
   \end{equation}

The solution of (40) is very simple, $ \Psi(x,y) = \psi(x) \varrho(y) $.
The $y$-part is just a harmonic oscillator\footnote{with a different
overall sign (relative to $H_{a}$). Remember that in Wheeler's
superspace, only the ``volume coordinate" (in our case, $a$)
is timelike, and all the other (infinite number of) dimensions are
spacelike.}, for which the WKB is the exact solution.
The frequency of this harmonic oscillator is  $ \omega = 16\pi\chi_{0}^{3}
$ . So the energy is $ \hbar \omega (m + 1/2) $ , and from (40) we get
   \begin{equation}
\left[ -\hbar^{2} \frac{d^{2}}{dx^{2}}+\tilde{V}_{m}(x) \right] \psi(x) = 0
   \end{equation}
where
   \begin{equation}
\tilde{V}_{m}(x) = {(8\pi \chi_{0}^{3})}^{2} \left[ x \left(x - a_{0} -
\frac{\alpha}{4\pi \chi_{0}^{3}} \right) - \frac{(m+1/2) \hbar}{4\pi
\chi_{0}^{3}} \right]
   \end{equation}
This is (22) and (23).

\vspace{2cm}
{\bf References}
\begin{enumerate}
\item J.R.Oppenheimer and H.Snyder, {\em Phys.Rev.}{\bf 56},455 (1939)\\
\hspace{0.6cm} C.W.Misner, K.S.Thorne and J.A.Wheeler, {\em Gravitation}
(Freeman, San Fransisco, 1973)
\item R.Arnowitt, S.Deser and C.W.Misner, {\em Ann.Phys.}(N.Y){\bf 33},
88 (1965)
\item F. Lund, {\em Phys.Rev.}{\bf D8}, 3253, 4229 (1973)
\item J.H.Kung, Harvard preprint (1993), hep-th/9302016
\item J.A.Wheeler, in {\em Battelle Recontres}, eds. C.M.DeWitt and J.A.
Wheeler (New York, Weily, 1968)
\item B.S.DeWitt, {\em Phys.Rev.}{\bf 160},1113,(1967)
\item S.W.Hawking, ``Quantum Cosmology", Lctures in Les-Houches summer
school, 1983.\\
\hspace{0.6cm} J.B.Hartle and S.W.Hawking, {\em Phys.Rev.}{\bf D28},
2960,(1983)\\
\hspace{0.6cm}J.B.Hartle,in {\em High Energy Physics}, eds. M.J.Bowick
and F.Gursey, (World Sceintific, 1986)
\item R.H.Price, {\em Phys.Rev.}{\bf D10},2419,2439 (1972)
\item U.H.Gerlach and U.K.Sengupta, {\em Phys.Rev.}{\bf D18},1773 (1978)
\item I.V.Novikov, Doctoral dissertation, Shternberg Astronomical
Institute, Moscow (1963)
\end{enumerate}

\end{document}